\documentclass[twoside,slac_one]{revtex4}
\usepackage{graphicx}
\usepackage{fancyhdr}
\usepackage{amsmath} % American Mathematics Society standards
\usepackage{bm}% bold math
\usepackage{amsxtra}
\usepackage{amssymb}
\usepackage{amsthm}
\usepackage{latexsym}
\usepackage{lscape}

\begin{document}

\title{Search for neutral Supersymmetric Higgs bosons in
di-$\tau$ and  ${b\tau\tau}$  final states in
${p\bar{p}}$ collisions at ${\sqrt{s}=1.96}$
TeV }


\author{Subhendu Chakrabarti \\for D0 and CDF collaborations}
\affiliation{Department of Physics and Astronomy, State University of New York, Stony Brook, NY, USA}

\begin{abstract}
We present a search for Higgs boson produced in  the di-tau modes or
via the associated $p\bar{p}\to h+b \to\tau^+\tau^-b$ process at a
center-of-mass energy of $\sqrt{s}=$1.96~TeV using up to 1.8-7.3~fb$^{-1}$ of data
collected with the D0 and CDF detector at the Fermilab Tevatron
collider. In Supersymmetric models Higgs boson production cross
section can be significantly enhanced compared to the Standard
Model. Additionally the Higgs boson has a significant branching
ratio to $\tau$ leptons at all masses.  The di-$\tau$ and $b-\tau\tau$
channels complement each other  providing best sensitivity for the search
in the SUSY parameter space.
\end{abstract}

\maketitle

\thispagestyle{fancy}


%%%%%%%%%%%%%%%%%%%%%%%%%%%%%%%%%%
\section{Introduction}

In the minimal supersymmetric standard model (MSSM), two complex Higgs bosons doublets can exist unlike only one standard model Higgs ~\cite{intro1}. These doublets lead to five physical Higgs bosons: two neutral CP-even (h,H), one neutral CP-odd (A) and two charged Higgs Bosons (H$^{\pm}$).  The three neutral Higgs bosons are collectively called as $\phi$. At the tree level, Higgs sector of the MSSM  are controlled by two parameters, which are the mass of the CP-odd Higgs bosons $M_{A}$, and the ratio of the vacuum expectation values of the two Higgs doublets, tan $\beta$. The neutral bosons decay into $\tau^{+}\tau{-}$ and $b\bar{b}$ pairs with branching fractions $10\%$ and  $90\%$.  The neutral bosons also can decay in association with b quarks which lead to complimentary search channels to inclusive production.  The production cross-section is enhanced by a factor that depends on tan $\beta$  than the standard model Higgs boson cross-section of the same  mass. Also, for large tan $\beta$,  the Higgs Boson A and either h or H are nearly degenerate in mass which leads to an approximate doubling of production cross-section.\\
In this paper, we present summary of MSSM neutral Higgs bosons searches  with the data collected by D0 and CDF detector, described in  ~\cite{det} at the Tevatron involving  $\tau$ leptons. These searches are complimentary to searches  with $b\bar{b}$ pairs which is leading decay mode. Despite lower branching ratio, searches with  $\tau^{+}\tau{-}$ pair have other advantages  that they are not suppressed with huge multijet and  large di-jet background. Moreover, neutral Higgs production in association with  b quarks has the advantage to use the information of  tagged b quarks to suppress irreducible $Z$+jets  background.

%%%%%%%%%%%%%%%%%%%%%%%%%%%%%%%%%%

\section{Inclusive Higgs Boson Searches }

In this section,  we briefly describe the search for neutral Higgs bosons  which  requires reconstruction of muons or electrons  and atleast one  hadronic decays of  $\tau$ lepton ($\tau_{had}$) and and missing transverse energy.  A details descrption can be found elsewhere ~\cite{d0incl}. The data considered in this analysis were recorded by the D0 detector and correspond to an integrated luminosity of 5.4 fb$^{-1}$ for the  channel. 
Electron are reconstructed using their characteristic energy deposits and the shower shape in the calorimeter.  Muons are identified from track segments in the muon system that are spatially matched to reconstructed in the muon system. Hadronic $\tau$ decays are reconstructed from energy deposits in the calorimeter using a jet cone algorithm. The $\tau$ candidates are then split into three categories which roughly correspond to one-prong $\tau$ decay with no $\pi^{0}$ (type 1), one-prong decay with  $\pi^{0}$  (type 2) and multiprong decay  (type 3). In additional, we use a neural-network based  $\tau$ identification to separate jets from real  $\tau$ production. The NN$_\tau$ is based on shower shape variables, isolation variables and correlation variables between tracking and calorimeter energy measurements. 

\begin{figure}[ht]
\centering
\includegraphics[height=50mm,width=60mm]{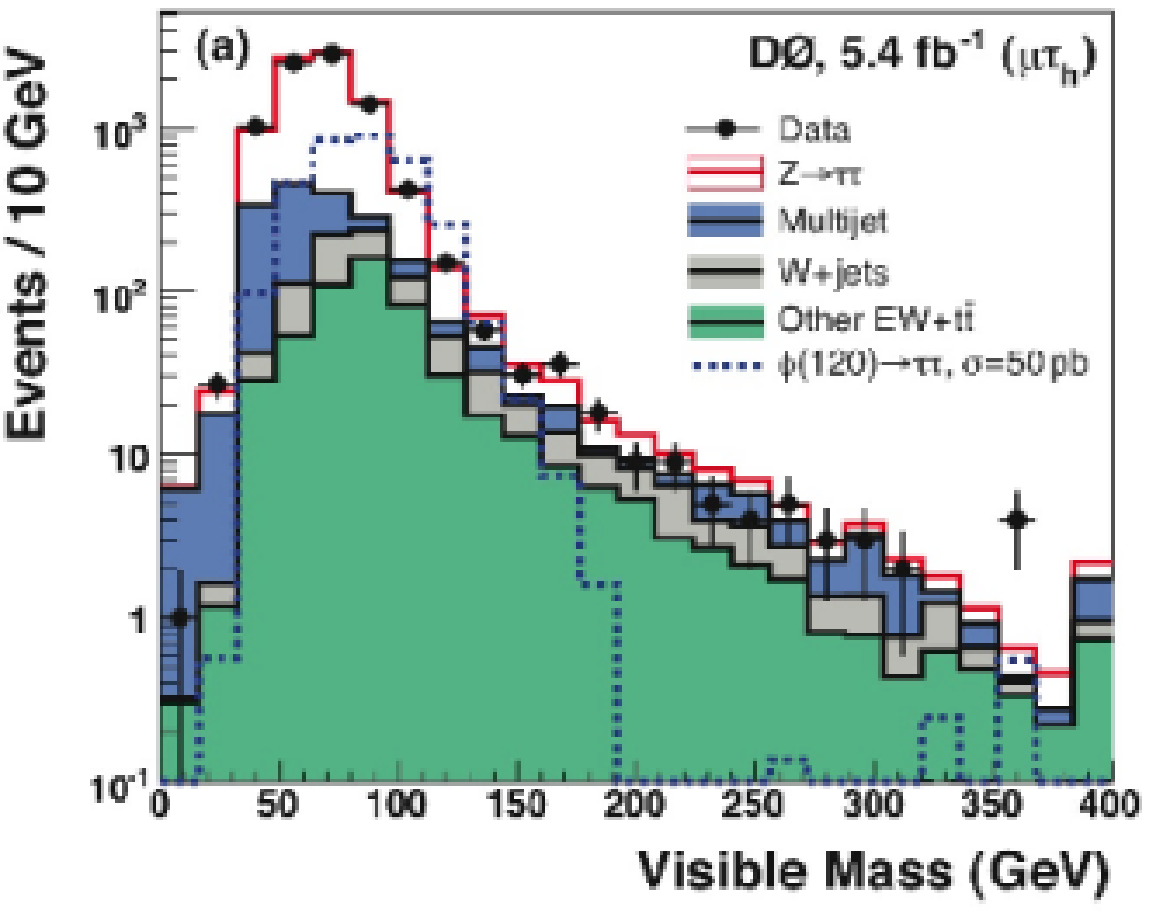}
\includegraphics[height=50mm,width=60mm]{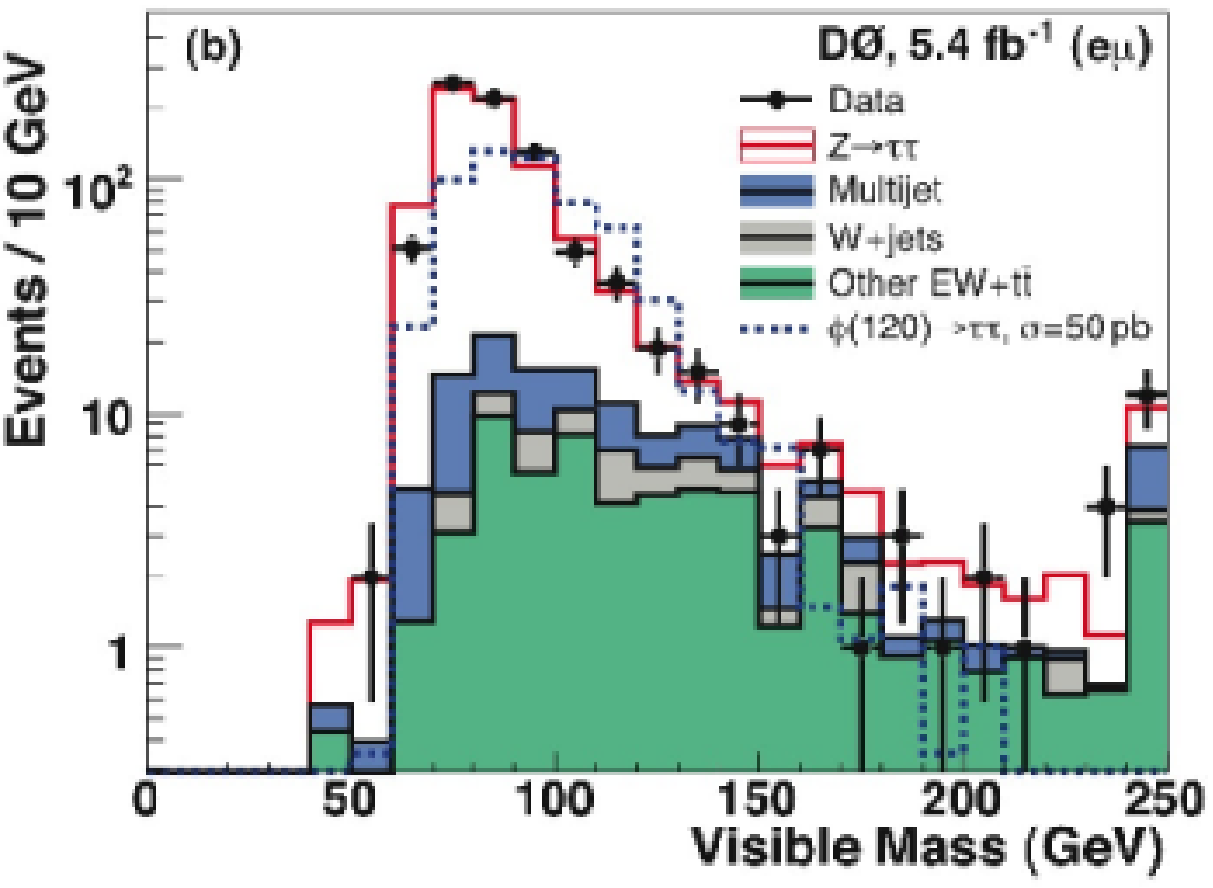}
\caption{ Visible mass distribution is shown.  $\mu \tau_{h}$  on left and $e\mu$ on right} \label{preseld0incl}
\end{figure}

Events are selected by requiring at least one single muon trigger for the $\mu \tau_{h}$ channel, while for the $e \mu$ channel, they need to fulfill either inclusive electron and muon trigger conditions.  The preselection sample is dominated by $Z$+jets, $W$+jets and MJ backgrounds. To reject  $Z\gamma \rightarrow \mu \mu$ and  $W$+jets background events,  further selection cuts  $\delta (\tau, \mu) >$ 0.5 and transeverse mass of muon and missing energy $>$ 50 are applied. In the $e\mu $ channel, events with atleast one muon with an oppositely electron are selected. Multijet background and W boson production is suppressed by requiring the mass of $e\mu$ pair to be larger than 20 GeV and sum of missing transverse energy and transverse momentum of electron and muon $>$ 65 GeV. Requiring the scalar sum of the transverse momenta of all jets to be $<$ 70 GeV rejects a large fraction of $t \bar{t}$ events.  Shape of multijet background is estimated from same-sign data reversing the NN cut on $\tau$s. For $e\mu$ channel, electron likelihood and muon isolation criteria is inverted to obtain the multijet background shape. After event selection, the number of expected SM background events (the number of observed data) are  8782 (8774), 830 (825) for the $\mu \tau_{h}$ channel and $e \mu$ channel.  The  signal  efficiency from these channels  is 1.16\% (0.2\%). After imposing all selection requirements, a variable called visible mass is reconstructed which is calculated using the four-vectors of the measured tau-lepton decay products. Distributions for the visible mass variable are shown in Fig.~\ref{preseld0incl}.

Several sources of systematics uncertainty affect the signal efficiency and background estimation. Largest systematics sources are integrated luminosity (6.1\%), muon identification (2.9\%), tau  identification (12\%, 4.2\%, 7\% per $\tau-$type), electron identification  (2.5\%), $Z+$jets cross-sections (5\%), $W+$jets cross-sections (10-20\%) and modeling of multijet background (9.1\%, 17.7\%, 12.5 per $\tau-$type).

CDF detector also performed this search for inclusive production of neutral Higgs bosons  using data correspond to   an integrated luminosity of 1.8 fb$^{-1}$ ~\cite{cdfincl}.  CDF considered  di$\tau$ pairs in three final states which   require reconstruction of electron, muons   and hadronic decays of  $\tau$ lepton ($\tau_{had}$). 

\begin{figure}[ht]
%\centering
\includegraphics[width=50mm]{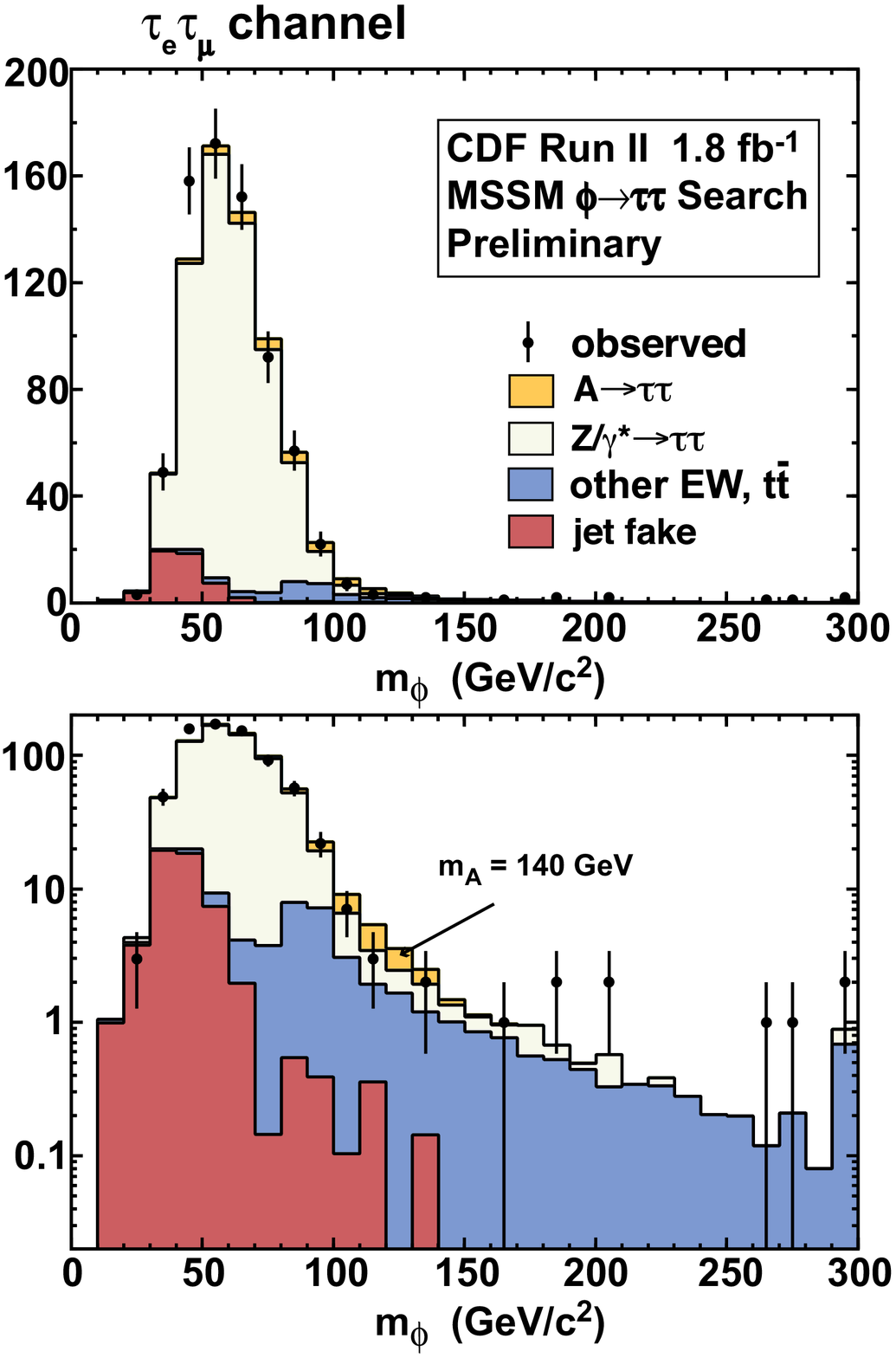}
\includegraphics[width=50mm]{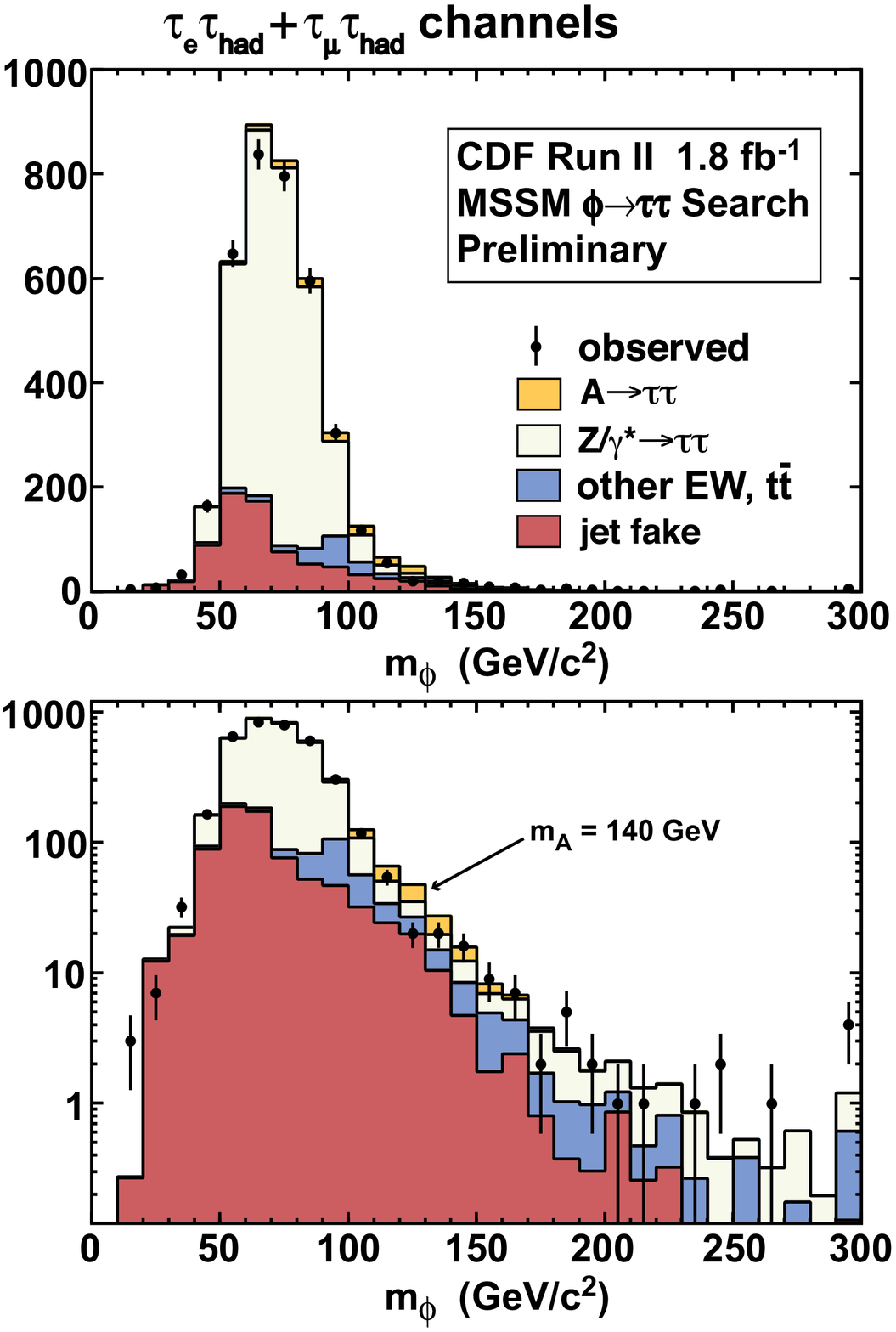}
\caption{ Visible mass distribution is shown.  $\mu \tau_{h}$, $e \tau_{h}$  on right and $e\mu$ on left} \label{preselcdfincl}
\end{figure}

The events for the  electron or muons   and atleast one hadronic decays of  $\tau$ required to have lepton plus track triggers. They also require a lepton candidate and an isolated track with tranverse momentum $p_T >$ 5 GeV, both pointing the central detector ($|\eta| \ge 1 $) and having azimuthal separtion $\delta \phi >$10 $^{0}$. In the CDF, the decay products in $\tau_{had}$ appear as narrow jets with low track and $\pi^{0}$ multiplicity. Reconstrcution of  $\tau_{had}$ use a seed track and contigous towers in the EM calorimeter for a tau candidate. The dominant background is  $Z$+jets which can come from a Z$\gamma$ production with subsequent decay to $\tau$ pair.  The second-largest contribution is from backgrounds from multijet or di-jet events where jets arising form quark or gluon fake as taus.  There are also another set of standard model backgrounds which include $Z$, $WW$, $WZ$, $ZZ$, $W\gamma$, Z$\gamma$ and $t\bar t$ production.  The number of expected SM background events (the number of observed data) are  1921.1 (1979), 1750.8 (1666) and 701.9 (726). The combined signal detection efficiency from  three channels  for a Higgs boson of mass 90 GeV (250) is 1.0\% (3.1\%). To probe for possible Higgs mass, we performed binned likelihood fits for the partial reconstructed  mass of di-$\tau$  (m$_{vis}$) as shown in  Fig.~\ref{preselcdfincl}. 

Several sources of systematics uncertainty also considered in this analysis.  Major contribution comes from multijet background estimation  (20\%,15\%),  for $\mu(e)\tau_{h}$ and $e \mu$ channels. Also PDF's introduce 5.7\%  systematics and  integrated luminosity contributes (6.1\%)  systematics +with minor contributions from electron and $\tau$ identifications.  

%%%%%%%%%%%%%%%%%%%%%%%%%%%%%%%%%%
\section{Higgs Boson Searches in association of a b quark}

In this section,  we briefly describe the search for neutral Higgs bosons  production in association with a b quark. The channel can produce electron or muons from leptonic $\tau$ decay, hadronic decays of  $\tau$ lepton ($\tau_{had}$) and a b quark jet ~\cite{d0btautau}.   The data considered in these analyses were recorded by the D0 detector  and correspond to an integrated luminosity of 3.7(7.3) fb$^{-1}$ for the electron (muon) channel respectively.

\begin{figure}[ht]
%\centering
\includegraphics[width=50mm]{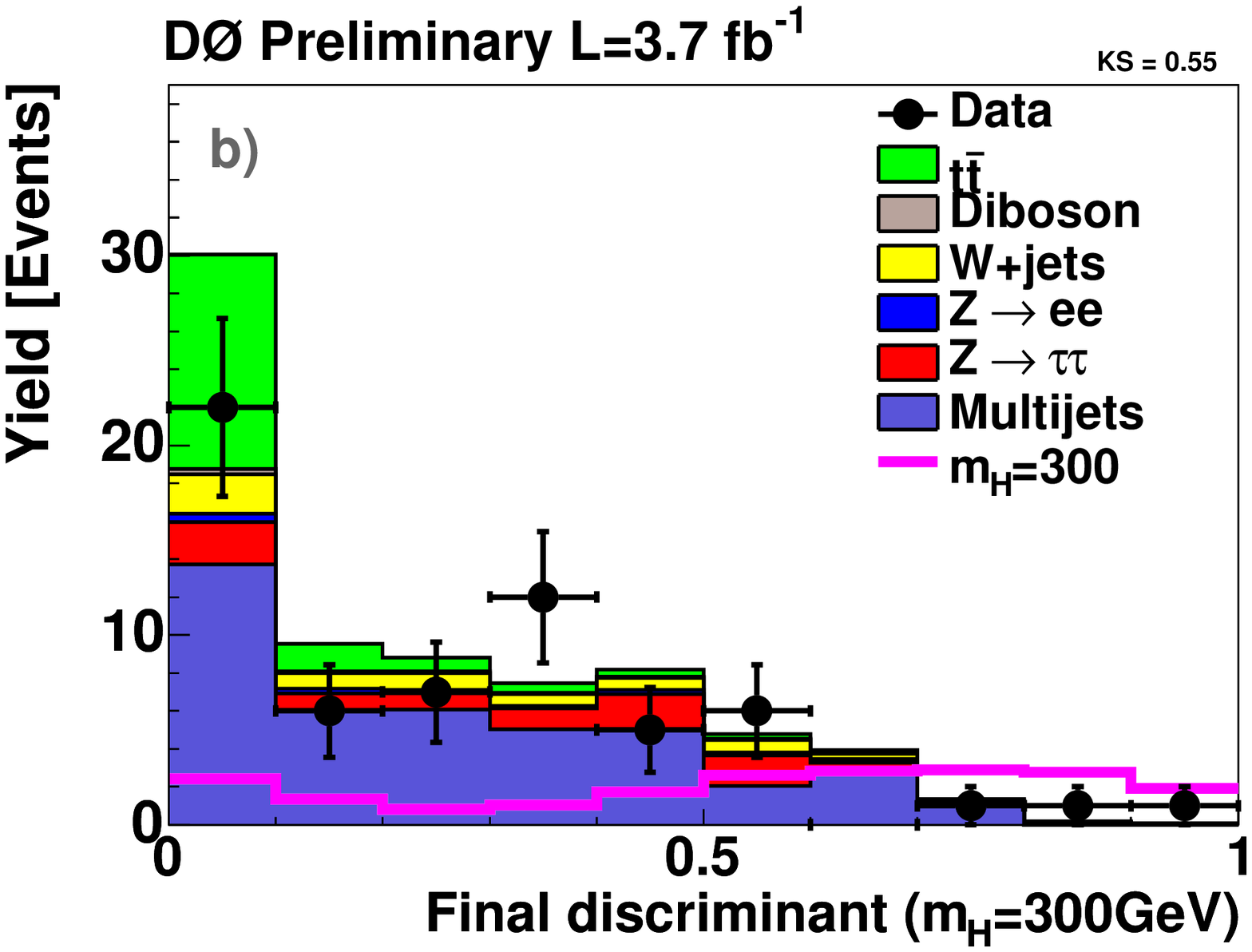}
\includegraphics[width=50mm]{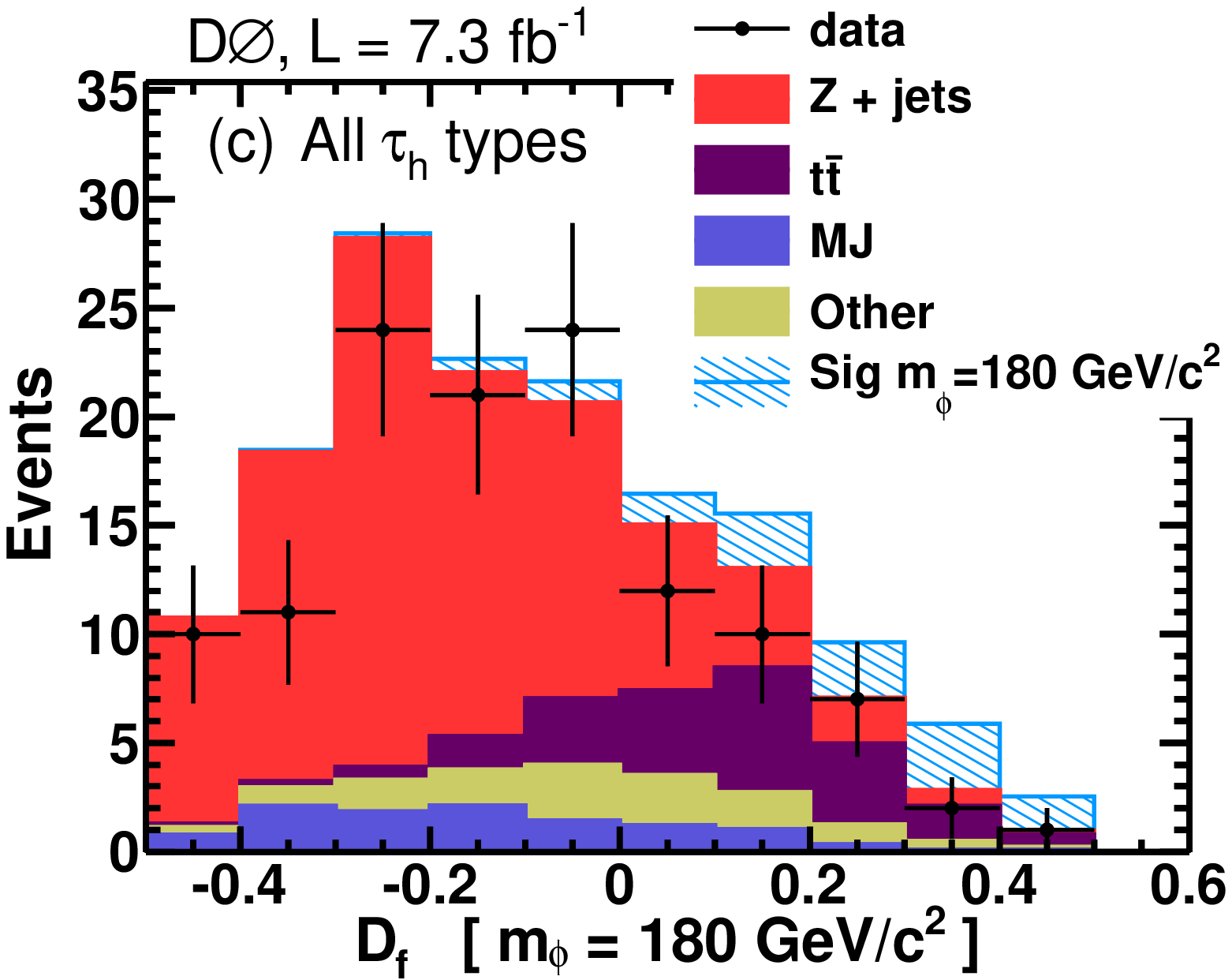}
\caption{ Final discriminant shape is shown for electron (left) and muon (right) channel. } \label{preselbtautaumu}
\end{figure}

The events are selected for the  electron or muons   and atleast one hadronic decays of  $\tau$  and one good jet. For electron channel, a neural-network to reject $t\bar{t}$, and  D$_{MJ}$ to reject MJ.  For muon channel,  most of the MJ background  is removed by the requirement D$_{MJ}>$0.1, where  D$_{MJ}$ is a multivariate discriminant described below. Finally, to improve the signal background ratio, we select a more restrictive b-tagged sample by demanding at least one jet to have $NN_b >$ 0.25. This b-tag requirement has an efficiency of 65\% for a probability of misidentifying a light jet as a b jet of 5\%. The number of expected SM background events (the number of observed data) are  2660.8(2629), 476.0(488)  for $e$ or $ \mu\tau_{had} b$ channel. For $ \mu\tau_{had} b$, a discriminant to reject $t\bar{t}$ and $M_{hat}$ variable is defined as the energy of $\tau_h \mu$ and its momentum along beam axis.  We use a multivariate technique with these variables in order to discriminate signal from background which is shown in   Fig.~\ref{preselbtautaumu} for both channels..
Eventually, the data are compatible with the background expectation for the final discriminant    and we proceed to set limits on production cross-section. Some of the  major systematics uncertainty contribution comes from MJ estimation 10-40\%, $Z$ + b-tagged jets normalizations(5\%), jet energy calibartion (10\%) and b-tagging (4\%) in addition to the uncertainty mentioned in above analysis. 

\section{Results}

\begin{figure}[ht]
%\centering
\includegraphics[width=50mm]{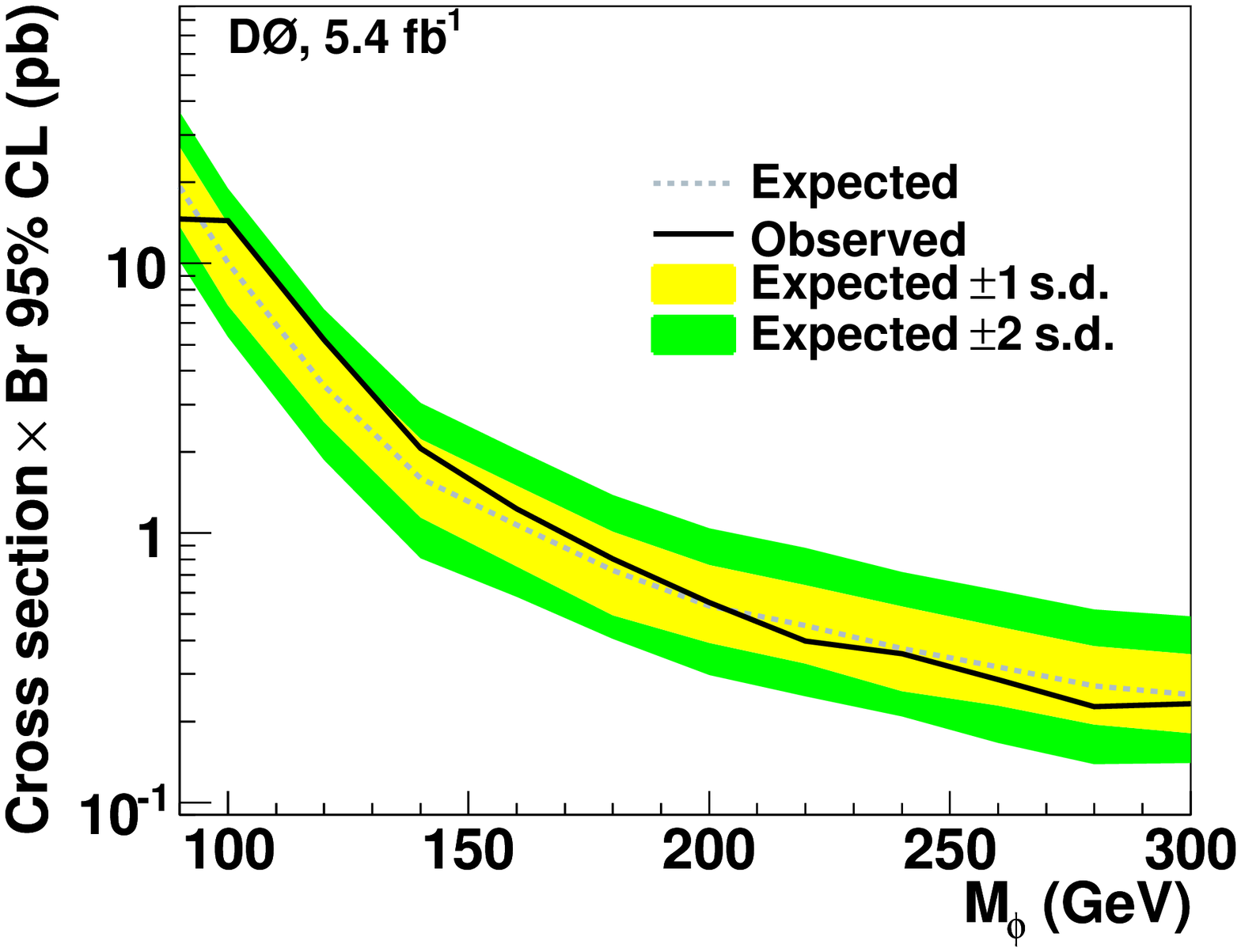}
\includegraphics[width=50mm]{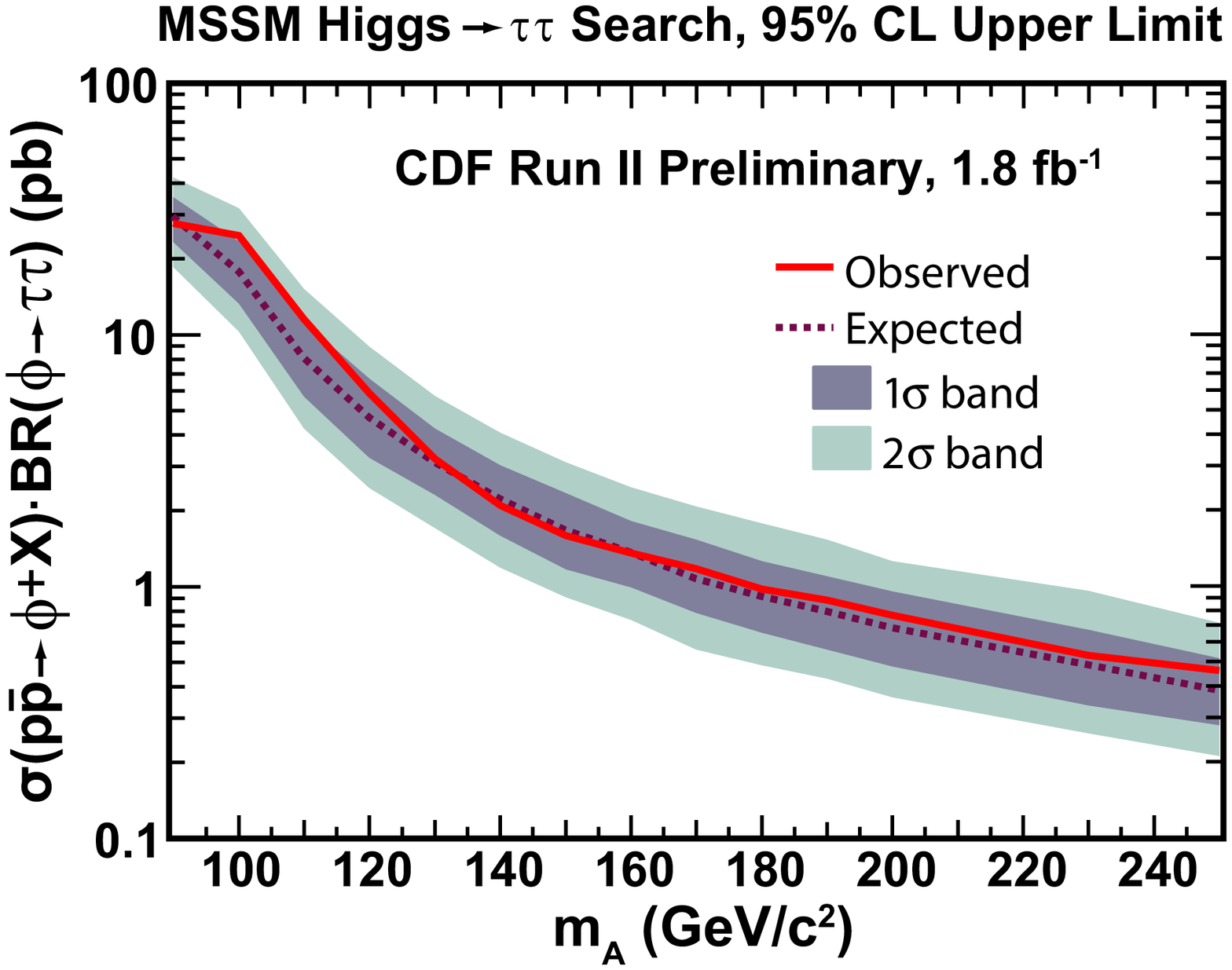}
\caption{Observed and expected limits at 95\% CL for Higgs production cross-section times branching ratio to $\tau$ pairs, CDF (right) and D0 (left)} \label{resd0cdfincl}
\end{figure}

\begin{figure}[ht]
%\centering
\includegraphics[width=50mm]{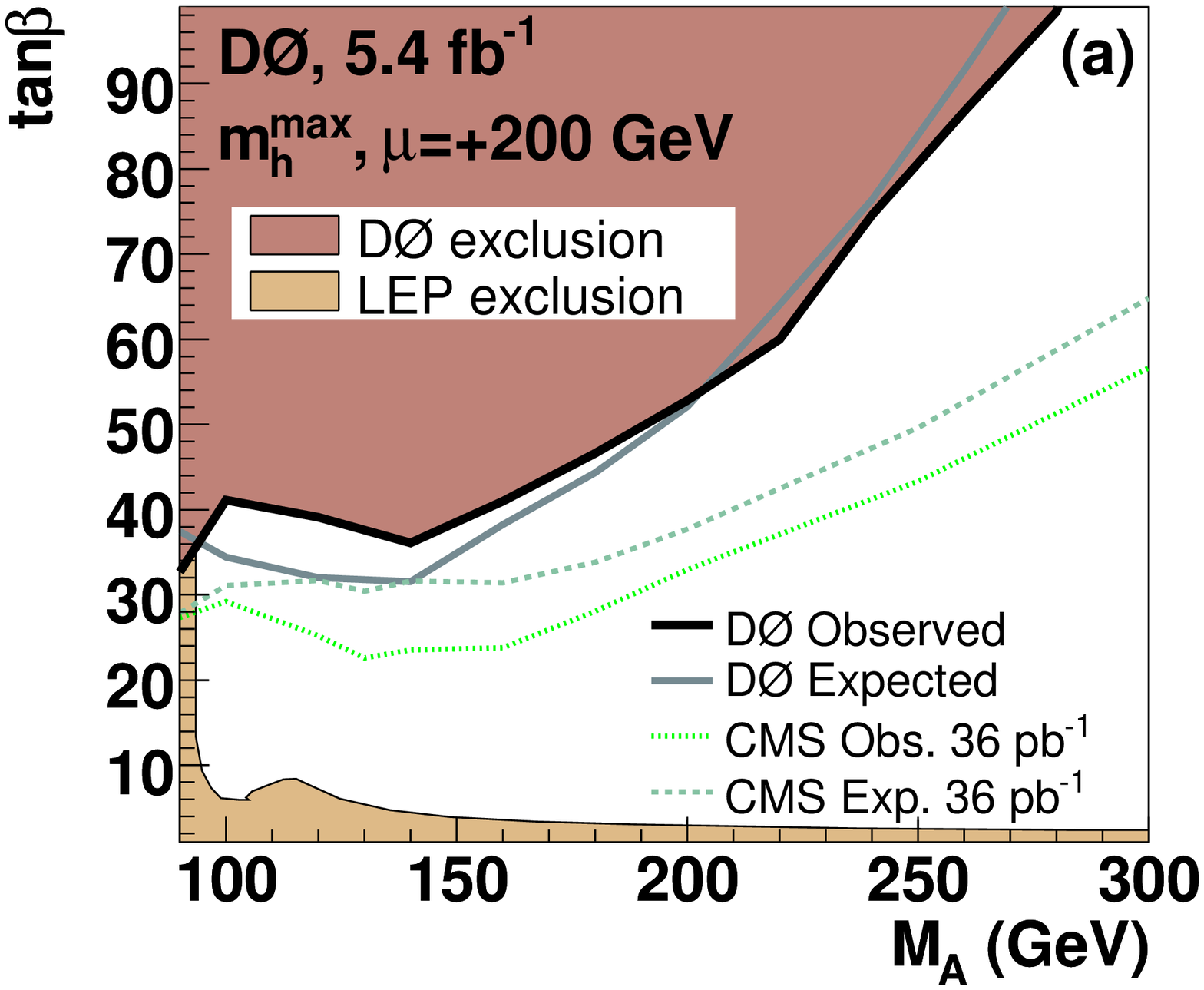}
\includegraphics[width=50mm]{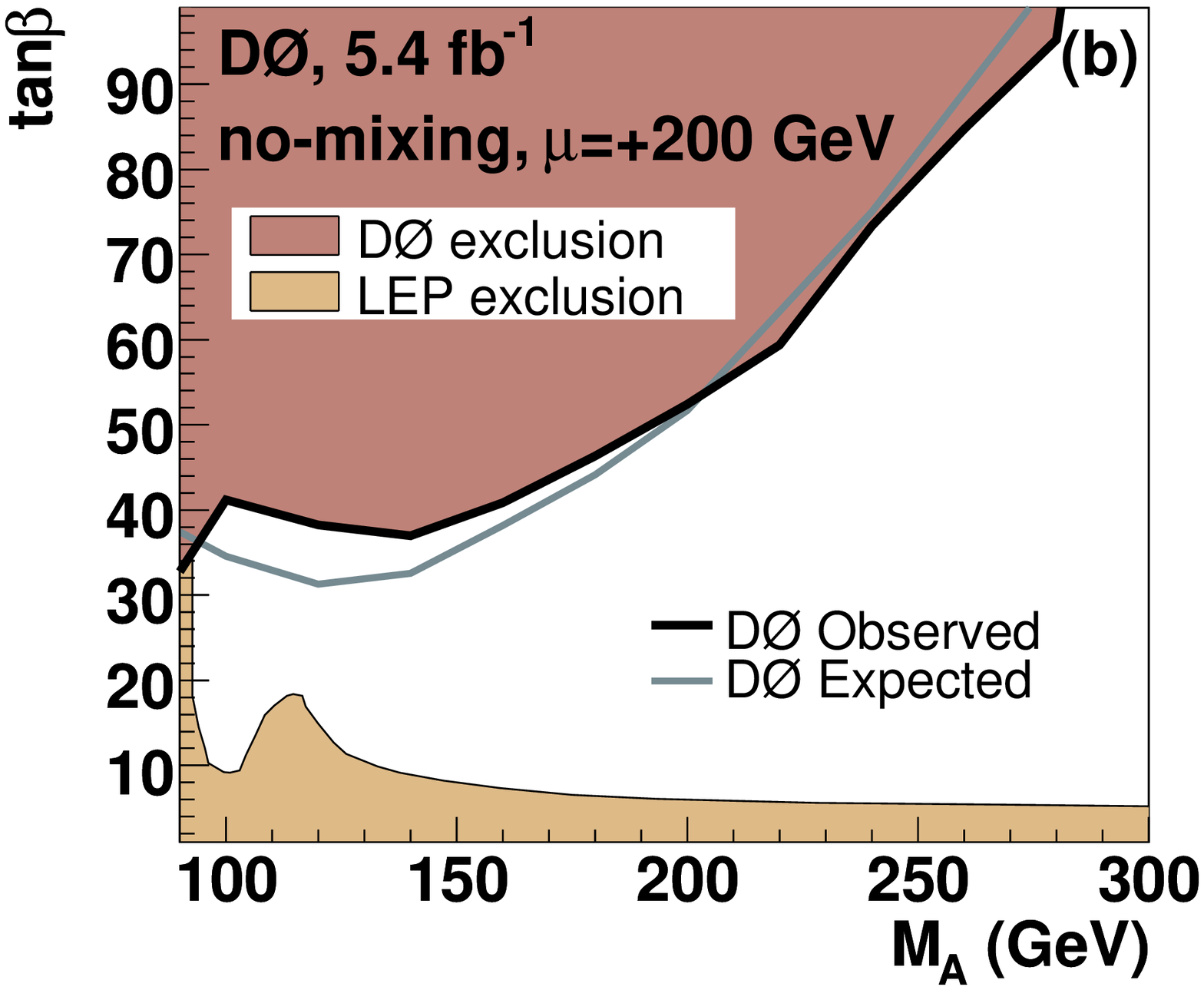}
\caption{Excluded region in tan$\beta$ vs $m_A$ plane for the ${m_h}^{max}$  with $\mu >$ 0 (left) and no-mixing scenarios with  $\mu <$ 0 (right) for D0} \label{limd0incl}
\end{figure}

\begin{figure}[ht]
%\centering
\includegraphics[width=50mm]{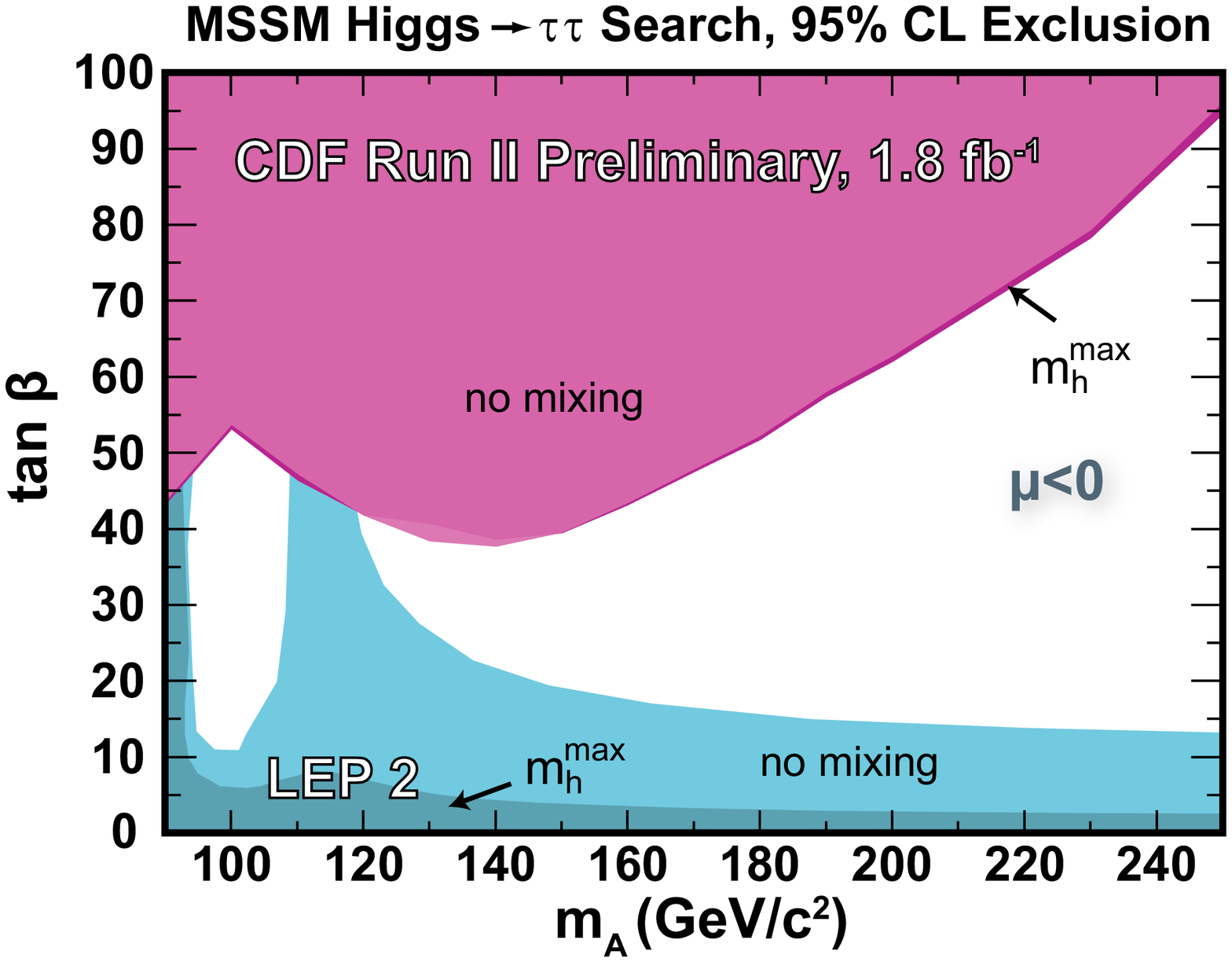}
\includegraphics[width=50mm]{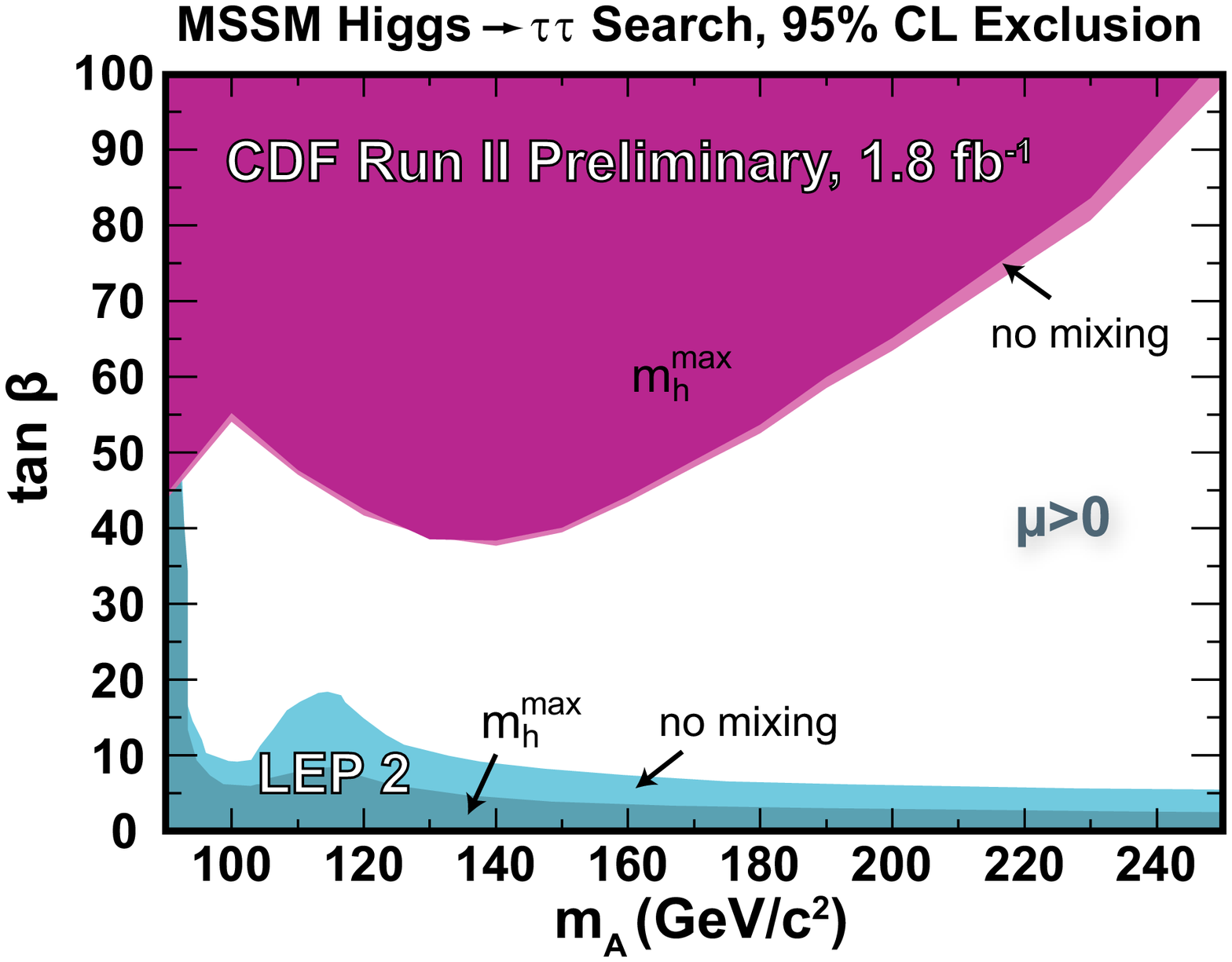}
\caption{Excluded region in tan$\beta$ vs $m_A$ plane for the ${m_h}^{max}$ and no-mixing scenarios with $\mu >$ 0 and  $\mu <$ 0 for CDF} \label{limcdfincl}
\end{figure}

\begin{figure}[ht]
%\centering
\includegraphics[width=45mm]{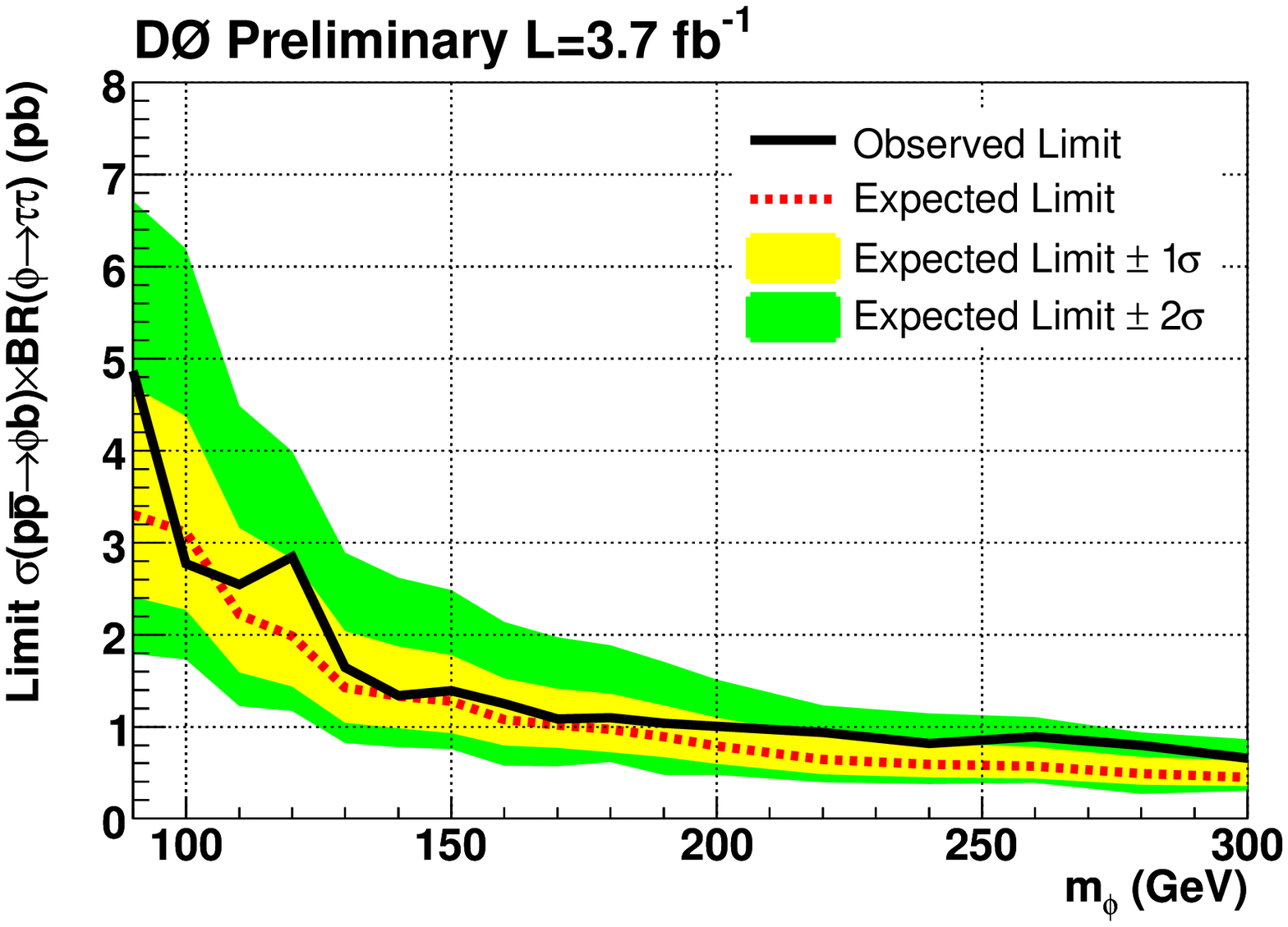}
\includegraphics[width=45mm]{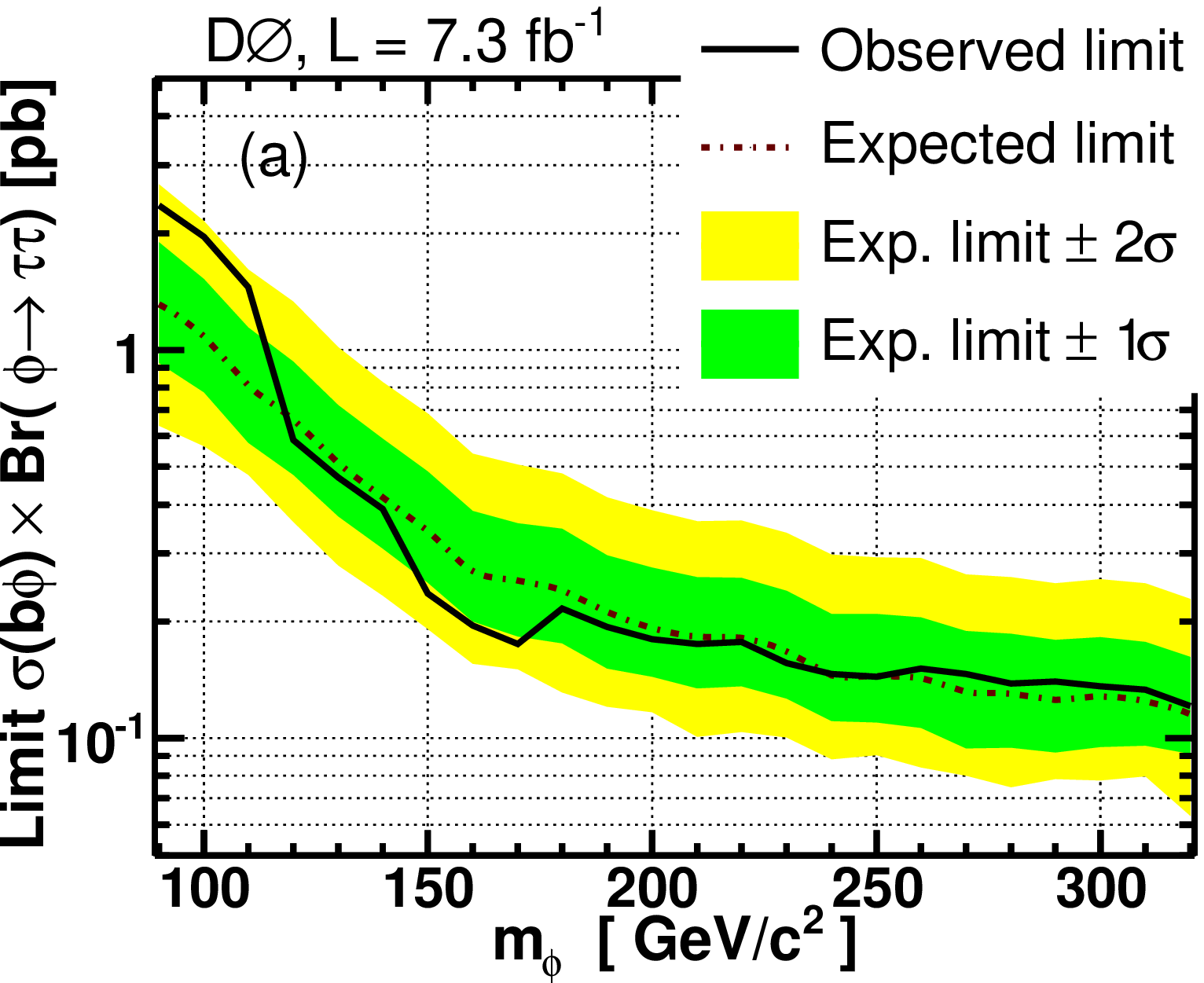}
\caption{Observed and expected limits at 95\% CL for Higgs production cross-section times branching ratio to $\tau$ pairs, muon (right) and electron channel at D0 (left)} \label{limbtautau}
\end{figure}

\begin{figure}[ht]
%\centering
\includegraphics[width=45mm]{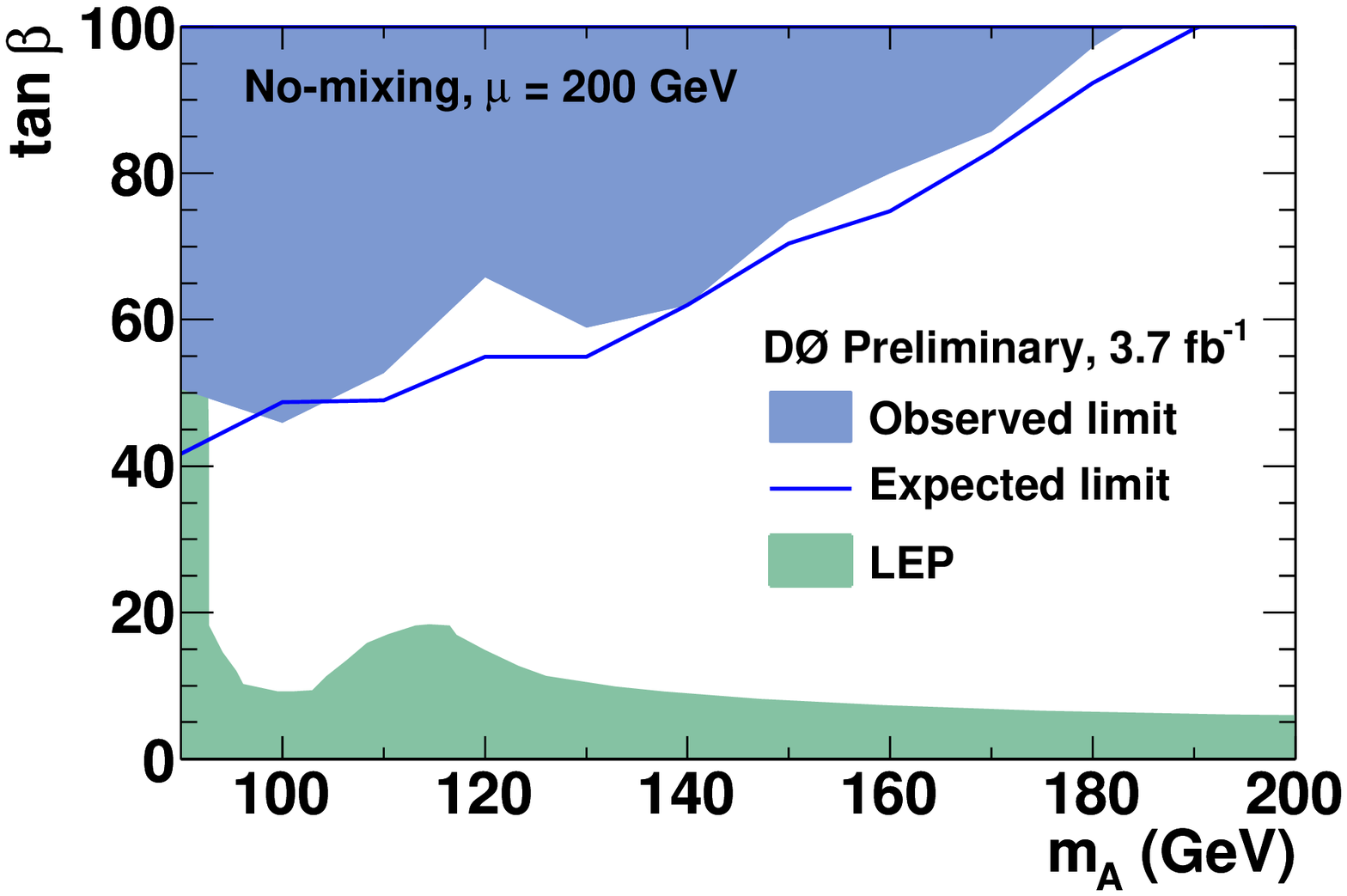}
\includegraphics[width=45mm]{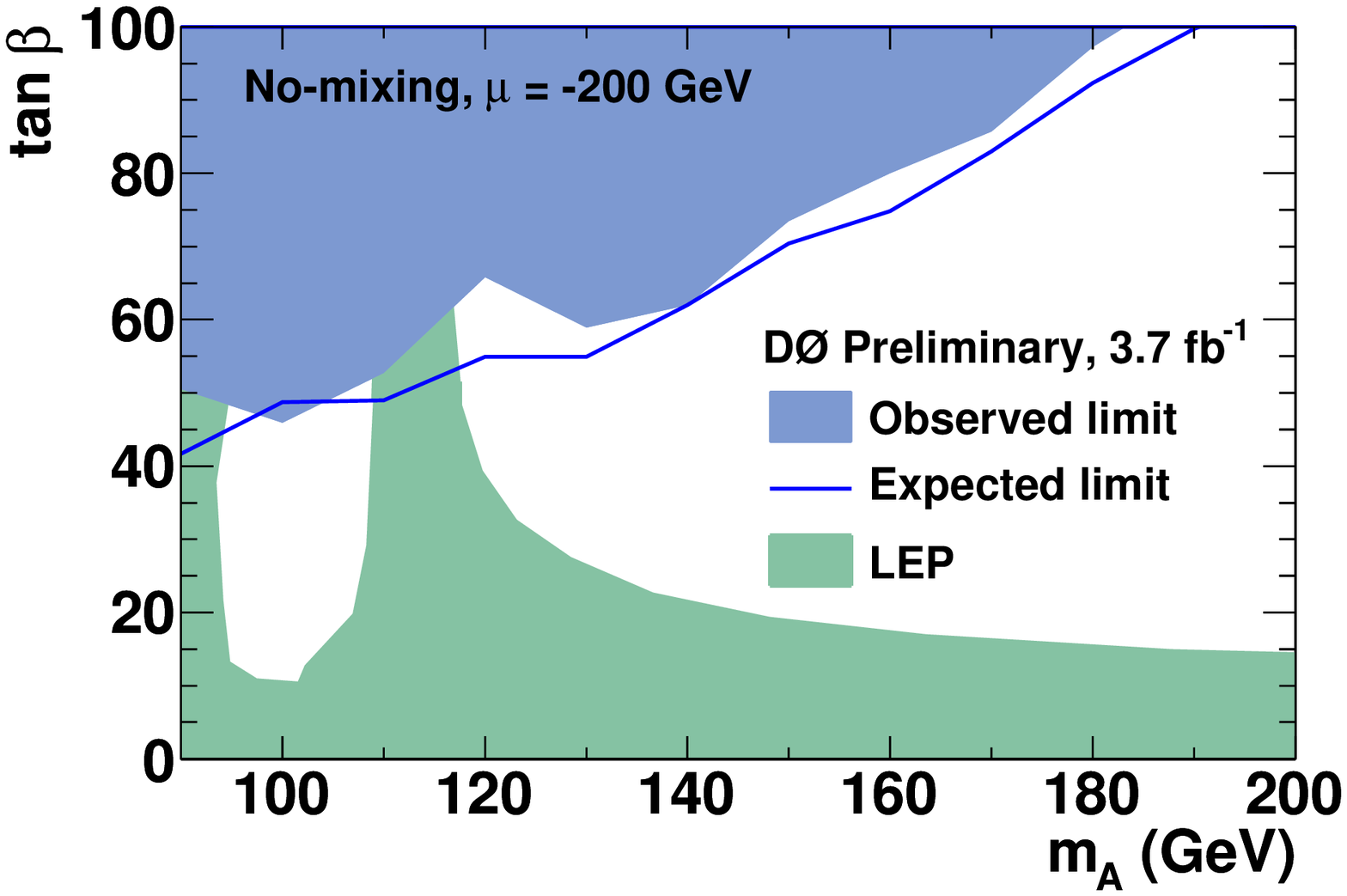}
\caption{Excluded region in tan$\beta$ vs $m_A$ plane for the ${m_h}^{max}$ and no-mixing scenarios with $\mu >$ 0 and  $\mu <$ 0 for D0 in electron channel} \label{resbtautauem}
\end{figure}

\begin{figure}[ht]
%\centering
\includegraphics[width=45mm]{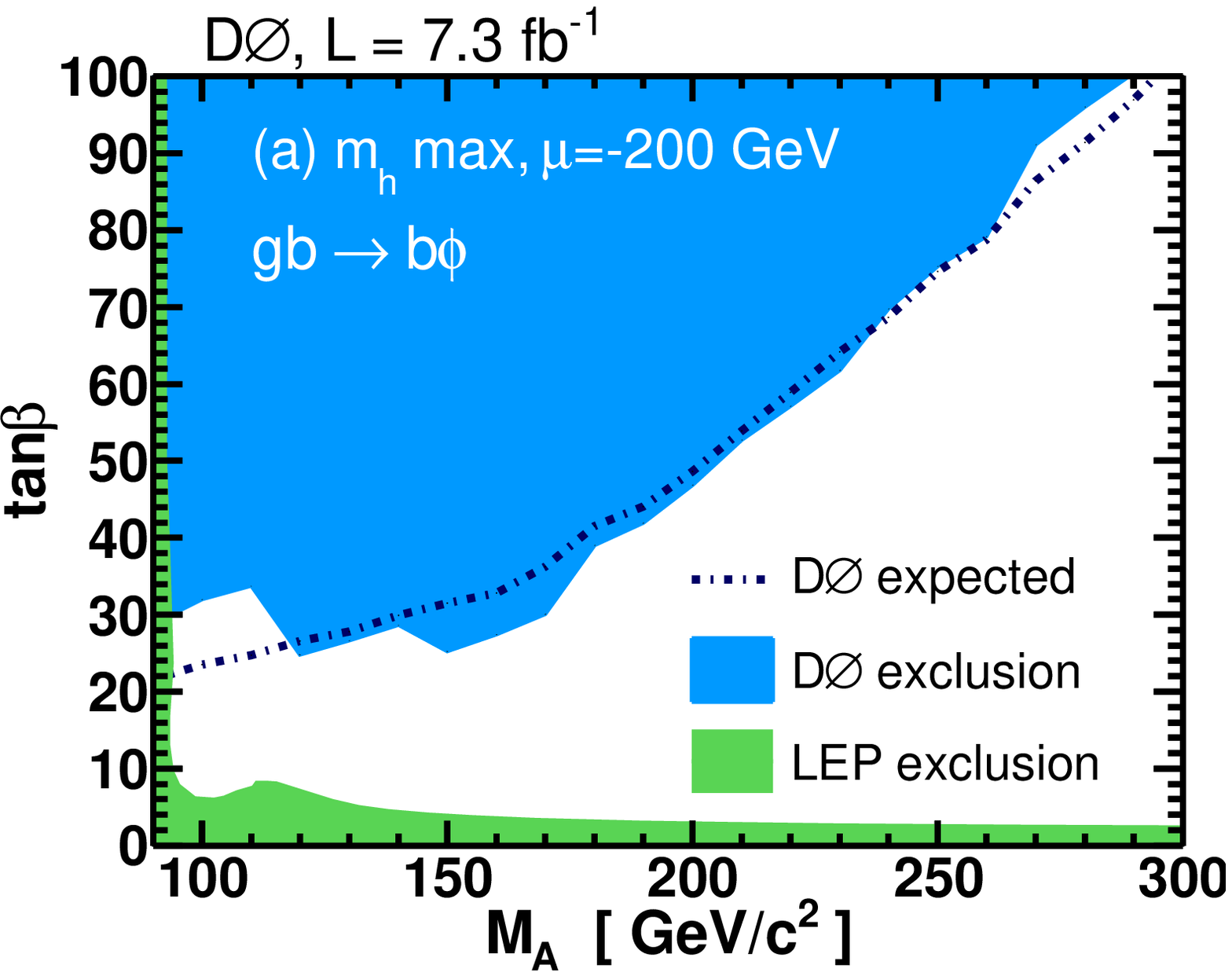}
\includegraphics[width=45mm]{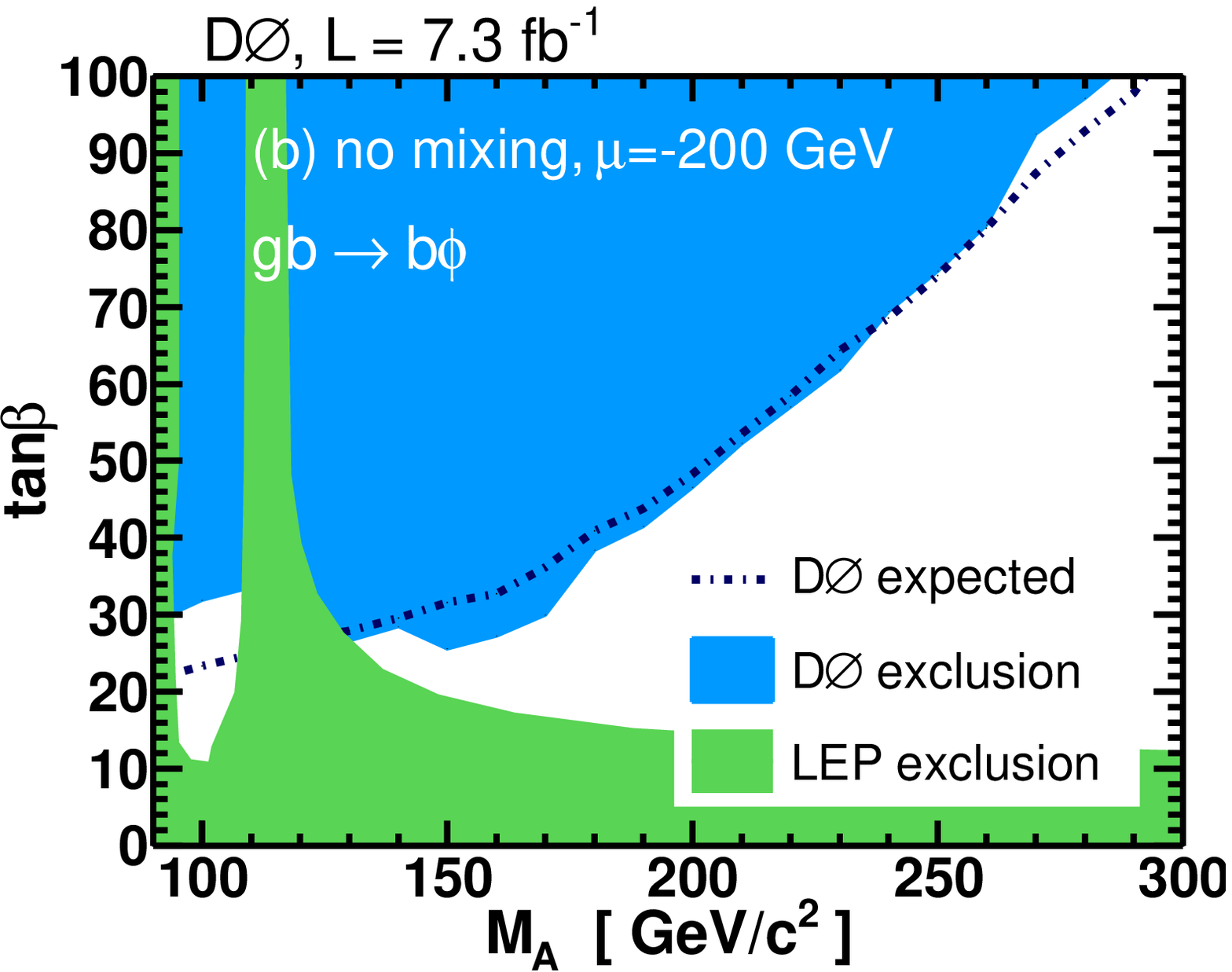}
\caption{Excluded region in tan$\beta$ vs $m_A$ plane for the no-mixing scenarios with $\mu >$ 0 (left) and  $\mu <$ 0 (right) for D0 in muon channel} \label{resbtautaumu}
\end{figure}

Visible mass distribution for the analysis is   used as input to a significance calculation using a modified frequentist approach with a Poisson log-likelihood ratio test statistic. In the absence of a significant signal, we derive upper limits for neutral Higgs boson  production cross-section and a profiling technique to reduce systematic uncertainties. The confidence level, CL$_s$ is defined as  CL$_s$ = CL$_{s+b}$ divided by  CL$_b$, where  CL$_{s+b}$ and  CL$_b$ are the confidence levels in the signal+background and background only hypothesis.  The combined limit on the mass of  neutral Higgs boson is set from  $m_A$ =90-250 GeV. An example fit in CDF inclusive search is also shown in   Fig.~\ref{resd0cdfincl}.    We observe no signal evidence in $m_A$ =90-250 GeV. The sensittivity assuming no signal is shown in and denoted as expected limits. Similarly,  limit on the mass of  neutral Higgs boson is set from  $m_A$ =90-320 GeV for muon channel (90-320 GeV for electron channel) with the complimentary channels as shown in  Fig.~\ref{limbtautau}. These limits  are translated into tan$\beta$, $m_A$ plane for two MSSM benchmark scenarios   the ${m_h}^{max}$ and no mixing scenarios with $\mu>$0  or $\mu<$0  and shown in Fig.~\ref{limd0incl},~\ref{limcdfincl} for  inclusive production. The exclusion plots in same parameter plane is shown in also Higgs production associated with a b quark electron and muon channel is shown Fig.~\ref{resbtautauem}, ~\ref{resbtautaumu}.   We exclude a substantial region of the MSSM parameter space, especially at low $m_A$ and set the most stringent limit to date at the Tevatron, involving $\tau$ final states.

%%%%%%%%%%%%%%%%%%%%%%%%%%%%%%%%%%

\bigskip % extra skip inserted

\end{document}